\documentclass[journal=jacsat,manuscript=article]{achemso}

\usepackage{chemformula} 
\usepackage[T1]{fontenc} 
\usepackage{float}
\usepackage{mhchem}

\author{Chunhao Bao}
\affiliation{International Quantum Academy, Shenzhen 518048, China.}
\author{Xiaolong Yin}
\affiliation{Department of Physics, Southern University of Science and
	Technology, Shenzhen 518055, China.}
\author{Jifeng Shao}
\affiliation{Shenzhen Institute for Quantum Science and Engineering, Southern University of Science and
	Technology, Shenzhen 518055, China.}
\alsoaffiliation{International Quantum Academy, Shenzhen 518048, China.}
\author{Longxiang Li}
\affiliation{Shenzhen Institute for Quantum Science and Engineering, Southern University of Science and
	Technology, Shenzhen 518055, China.}
\author{Zhiyue Li}
\affiliation{Shenzhen Institute for Quantum Science and Engineering, Southern University of Science and
	Technology, Shenzhen 518055, China.}
\alsoaffiliation{Department of Physics, The University of of Hong Kong, Hong Kong, 999077, China.}
\author{Xiaoming Ma}
\affiliation{Shenzhen Institute for Quantum Science and Engineering, Southern University of Science and
	Technology, Shenzhen 518055, China.}
\alsoaffiliation{International Quantum Academy, Shenzhen 518048, China.}
\author{Shu Guo}
\affiliation{Shenzhen Institute for Quantum Science and Engineering, Southern University of Science and
	Technology, Shenzhen 518055, China.}
\alsoaffiliation{International Quantum Academy, Shenzhen 518048, China.}
\author{Tingyong Chen}
\affiliation{Shenzhen Institute for Quantum Science and Engineering, Southern University of Science and
	Technology, Shenzhen 518055, China.}
\alsoaffiliation{International Quantum Academy, Shenzhen 518048, China.}
\email{chenty@sustech.edu.cn}

\title[An \textsf{achemso} demo]
  {Observation of Complete Orbital Two-channel Kondo Effect in van der Waals Ferromagnet \ce{Fe_{3}GaTe_{2}}}
\abbreviations{IR,NMR,UV}

\begin{document}
\bibliographystyle{unsrt}

\begin{abstract}
  
  Orbital two-channel Kondo (2CK) effect is one of the crucial systems with non-Fermi liquid (NFL) behaviors. But the full three-regime transport evidence has never been observed in one sample. Here, all three-resistive regimes for the orbital 2CK effect induced by two-level systems (TLSs) have been observed in the van der Waals ferromagnet \ce{Fe_3GaTe_2}. The electron behavior undergoes a continuous transition from electron scattering to the NFL behavior, and subsequently to Fermi liquid behavior. The magnetic field does not affect any regimes, indicating the non-magnetic origin of the TLSs in \ce{Fe_3GaTe_2}. In addition, the slope of linear negative magnetoresistance, rather than the topological Hall effect, has been found to be related to spin-magnon scattering and can be used to infer the emergence of spin textures. Our findings indicate \ce{Fe_3GaTe_2} may be an ideal platform to study electron-correlation and topological phenomena.
 
\end{abstract}

\section{Introduction}

The metallic state of many solids can be understood in the framework of Landau's Fermi liquid theory (FLT), where dynamics at low excitation energies and temperatures is described by substituting the non-interacting fermions with interacting quasiparticles carrying the same spin, charge and momentum$\cite{landau_theory_1956}$.  However, some of the most intriguing phenomena in strongly correlated systems lie beyond the quasiparticle paradigm where quantum criticalities may provide a better understanding$\cite{lee_recent_2018}$. Examples of such non-Fermi liquids (NFL) include Luttinger liquids$\cite{luttinger_exactly_1963,ishii_direct_2003}$, fractional quantum Hall Laughlin liquids$\cite{laughlin_anomalous_1983}$, high-temperature superconductors$\cite{laughlin_relationship_1988}$, heavy fermions$\cite{beri_topological_2012,cox_quadrupolar_1987}$ and the two-channel Kondo (2CK) system$\cite{potok_observation_2007}$. Although the Kondo ground state is complex, its excitations can still be described by FLT. In spin 2CK effect, however, a spin-1/2 impurity couples to conduction electrons into two equal orbital channels and leads to impurity quantum criticality with exotic NFL behavior as the consequence of two spins attempting to compensate the spin-1/2 impurity. But the spin 2CK effect is difficult to observe because of the strict requirements of zero local magnetic field and channel symmetry. Instead, an analogous orbital 2CK effect was proposed based on resonant scattering centers with orbital degrees of freedom such as two-level systems (TLSs)$\cite{zawadowski_kondo-like_1980,ralph_observations_1992,ralph_2-channel_1994}$.  In this scenario, the TLS assumes a role equivalent to that of the impurity spin-1/2 in the spin 2CK effect and is thus termed a pseudo-spin-1/2$\cite{zawadowski_kondo-like_1980,muramatsu_low-temperature_1986}$. Due to the larger orbital degree of freedom inherent in the TLSs, the orbital 2CK effect is supposed to be more readily observable. 

The 2CK effect has attracted significant attention due to its relevance to high-temperature superconductivity$\cite{landau_theory_1956}$, Majorana fermions$\cite{lee_recent_2018}$, and strongly correlated physics$\cite{luttinger_exactly_1963}$. The hallmark feature of the orbital 2CK effect is the three-resistive regimes in upturn resistance at low temperature: (1) Kondo regime: $T_\mathrm{K} < T < T_\mathrm{0}$, $R_\mathrm{xx} \sim$ $-$ln($T$), where the weak coupling starts at $T_\mathrm{0}$ and ends at Kondo Temperature ($T_\mathrm{K}$) due to electron scattering with the TLSs. (2) NFL regime: $T_\mathrm{D} (= \Delta^2/T_\mathrm{K}) < T \ll T_\mathrm{K}$, $R_\mathrm{xx} \sim$ $-T^{1/2}$, electrons compete to screen the pseudo-spin-1/2 and contribute to the “overscreened” of the scattering center$\cite{cox_exotic_1998,von_delft_2-channel_1998}$. This regime is absent in the Kondo effect and $\Delta$ is the energy splitting between the localized states. (3) Fermi liquid (FL) regime: $T < T_\mathrm{D}$, $R_\mathrm{xx} \sim -T^2$, where FL state shows up due to complete screening of the pseudo-spin-1/2. Because of the nonmagnetic origin of the TLSs, the three-resistive regimes do not depend on external magnetic fields. 

Previously, the resistance upturn has been observed in many systems, including Cu point contact$\cite{ralph_observations_1992,keijsers_two-level-system-related_1995,keijsers_point-contact_1996}$, glasslike ThAsSe$\cite{cichorek_two-channel_2005}$, epitaxial ferromagnetic $L$$1_0$-MnAl films$\cite{zhu_orbital_2016}$ and layered compound $\ce{ZrAs_{1.58}Se_{0.39}}$$\cite{cichorek_two-channel_2016}$. In addition, van der Waals (vdWs) ferromagnet such as $\ce{Fe_3GeTe_2}$ (FGeT) also show the characteristic of resistance upturn at low temperature$\cite{deng_gate-tunable_2018,fei_two-dimensional_2018}$. These effects are attributed to the orbital 2CK effect but not without controversies$\cite{yeh_two-channel_2009,schlottmann_multichannel_1993,zawadowski_kondo-like_1980,cox_exotic_1998}$, because the unambiguously three-resistive regimes expected from the orbital 2CK effect are not fully observed in a single material sample. Furthermore, NFL phenomena in vdWs materials are expected to deviate strongly from FL$\cite{jiang_non-fermi-liquid_2013}$. Ferromagnetic $\mathrm{Fe_3GeTe_2}$ (FGaT) is a unique vdWs material which has a Curie temperature ($T_\mathrm{C}$) of 340 K, above room temperature$\cite{zhang_above-room-temperature_2022}$. Since ferromagnetic thin films, such as $L$$1_0$-MnAl$\cite{zhu_orbital_2016}$, $L$$1_0$-MnGa$\cite{zhu_observation_2016}$ and FGeT$\cite{feng_resistance_2022}$ have already shown electrical transport characteristics associated with the orbital 2CK effect, FGaT may represent an ideal vdWs material platform to study the orbital 2CK effect. 

In this paper, we have observed the hallmark feature of the three-resistive regimes of temperature-dependence in a single FGaT material system, unambiguous evidence for the orbital 2CK effect. The longitudinal resistance $R_\mathrm{xx}$ undergoes a transformation across three consecutive temperature regimes, from electron-TLS scattering ($\sim$ $-$ln($T$): 30 K - 9 K) to NFL behavior ($\sim$ $-T^{1/2}$: 9 K - 1 K), then finally the FL behavior ($\sim -T^2$: < 1 K) due to complete screening of the pseudo-spin-1/2. The fact that magnetic fields up to 9 T do not disrupt the three-regimes behavior indicates that the orbital 2CK effect in FGaT originates from nonmagnetic TLSs. Topological Hall effect is not observed and the antisymmetric Hall peaks at room temperature are due to the anomalous Hall effect. We have found out that the linear negative magnetoresistance (LNMR) can be useful to characterize spin textures in magnetic materials. For decreasing temperature, the slope of the LNMR of the FGaT sample does not decrease monotonically due to the increase of the spin-magnon scattering around 130 K. 

\section{Results}

\begin{figure}[H]
	\centering
	\includegraphics{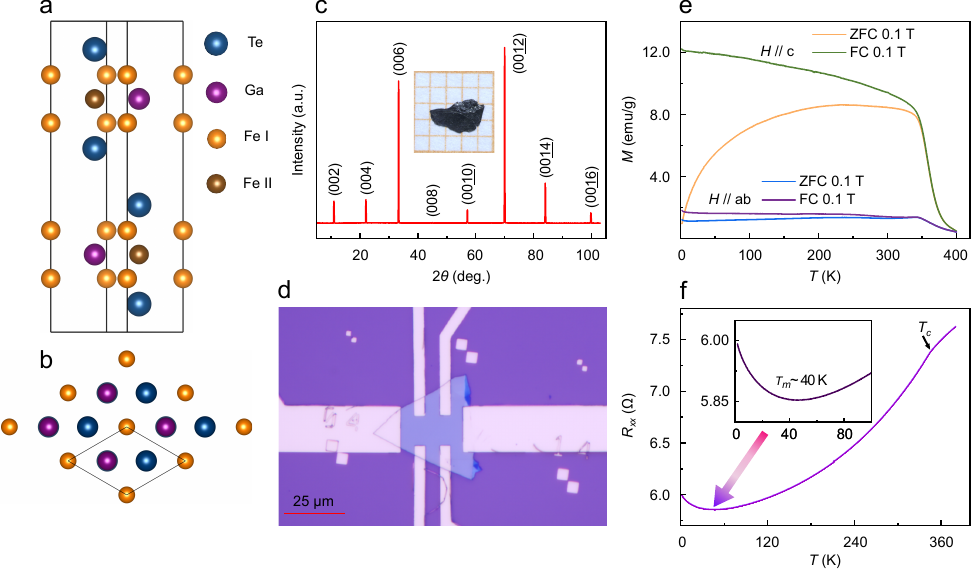}
	\caption{Crystal structure and transport characterization of thin  FGaT flakes. (a) Atomic structure of a FGaT bilayer. Fe(I) and Fe(II) indicate the two distinct Fe sites with +3 and 0 valence states. (b) Top view of FGaT. (c) XRD pattern of newly cleaved single crystal FGaT. The inset shows the optical image of FGaT single crystal. (d) Optical image of FGaT nanoflake Hall bar for transport measurement. Sample thickness: 81.4 nm. (e) Temperature-dependence of magnetization during zero-field-cooling (ZFC) and field-cooling (FC) process for  $H$$\parallel$$c$ and $H$$\parallel$$ab$ at $\mu_0H$ = 0.1 T. (f) Temperature dependence of longitudinal resistance $R_{xx}$. The inset shows the resistance upturn around 40 K. }
	\label{figure 1}
\end{figure}

The ferromagnet FGaT single crystals are synthesized by self-flux method and single crystal X-ray diffraction (XRD) shows a hexagonal structure with a space group $\mathrm{P6_3/mmc}$ ($a$ = $b$ = 4.0767 Å, $c$ = 16.088 Å).  In each layer of FGaT, covalently bonded $\mathrm{Fe_3Ga}$ contains the hexagonal Fe(I)-Ga atomic ring layer and two separated triangular Fe(I)-Fe(I) lattice layers, and both are sandwiched by two adjacent Te atomic layers with weak vdWs interlayer coupling, as shown in Figure \ref{figure 1}a, \ref{figure 1}b. The XRD pattern of the FGaT single crystal in Fig. \ref{figure 1}c with all (00l) Bragg peaks indicates that the $c$-axis is perpendicular to the newly cleaved surface of $ab$-plane. Following mechanical exfoliation, the FGaT thin films are transferred onto substrates and further patterned as the transport measurement structure, as illustrated in Fig.  \ref{figure 1}d. Fig.  \ref{figure 1}e is the temperature-dependent magnetization of the FGaT single crystal. These measurements were conducted under conditions of ZFC and FC, with a magnetic field of 0.1 T applied along both the $c$-axis and the $ab$-plane separately. The Curie temperature ($T_\mathrm{C}$) is around 340 K. The splitting of ZFC and FC curves indicates the formation of multidomain at low temperature. The field-cooled magnetization measured with $H$$\parallel$$c$ is much larger than that of $H$$\parallel$$ab$, revealing the out-of-plane easy axis of the FGaT crystal flake. Fig. \ref{figure 1}f shows the temperature-dependence of in-plane longitudinal resistance $R_{xx}$ cooling from 380 K to 1.8 K at $\mu_0H$ = 0 T with a current in the $ab$-plane. The resistance curve shows a clear kink at $T_\mathrm{C}$ around 340 K.  Subsequently, the minimum longitudinal resistance emerges around 40 K, after which the resistance shows an abnormal upturn as the temperature decreases. This is a potential sign for the orbital two-channel Kondo (2CK) effect in FGaT.

\begin{figure}[H]
	\centering
	\includegraphics{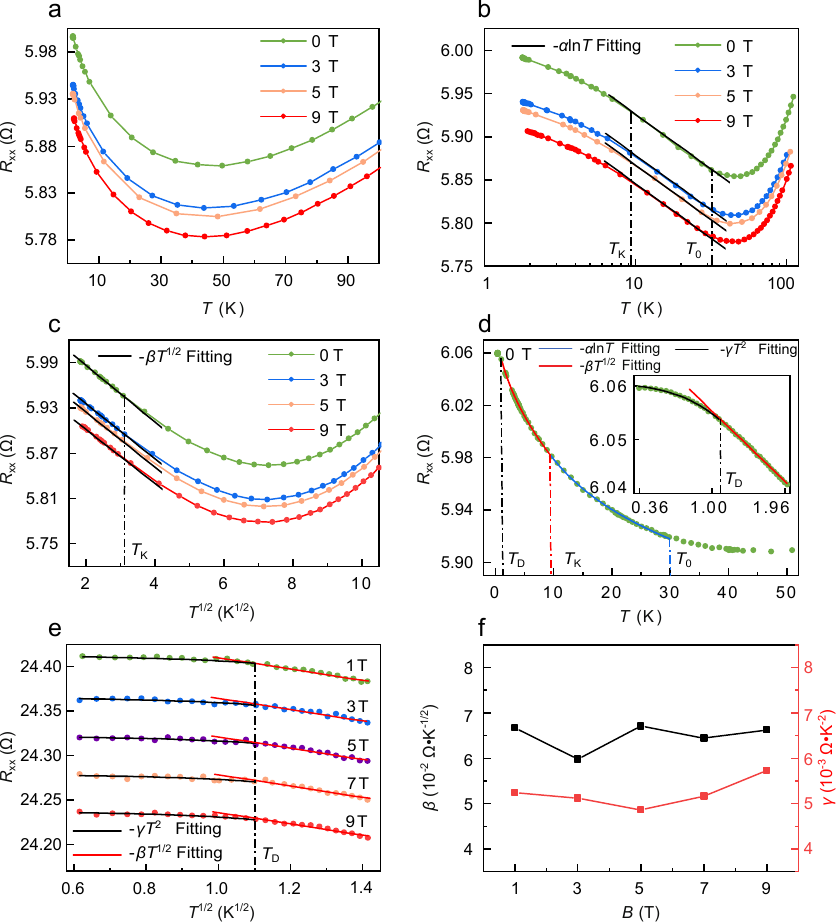}
	\caption{Evidence for orbital 2CK effect and Fermi liquid behavior down to 0.36 K. (a) The longitudinal resistance $R_{xx}$ vs. $T$ at different magnetic fields. (b) Semilog plot of the longitudinal resistance $R_{xx}$ vs. ln($T$) under different fields. The dotted lines are experimental data. (c) The longitudinal resistance $R_{xx}$ vs. $T^{1/2}$ under different fields. The dotted lines are experimental data. (d) Temperature-dependence $R_{xx}$ from 50 K to 0.36 K under no external field. The solid dots are the experimental data. The inset shows the longitudinal resistance $R_{xx}$ from 2 K to 0.36 K at $\mu_0H$ = 0 T. (e) The longitudinal resistance $R_{xx}$ vs. $T^{1/2}$ from 2 K to 0.36 K under different magnetic fields after the sample was oxidized in atmosphere for 7 days. The dotted lines are experimental data. (f) The variation of the slope parameters $\beta$ and $\gamma$ with the magnetic field.}
	\label{figure 2}
\end{figure}

To further reveal the mechanism behind the resistance upturn in FGaT, we measured the temperature-dependence of the longitudinal resistance under different magnetic fields up to 9 T. As shown in Fig. \ref{figure 2}a, the resistance minimum remains at the same temperature in the magnetic field, and the resistance upturn exhibits no sign of change except a vertical shift with magnetic fields up to 9 T. The persistence of resistance upturn under a magnetic field suggests that the weak localization effect will not be responsible for the resistance upturn in the FGaT system$\cite{niu_evidence_2016}$. To characterize the evolution of temperature-dependent resistance, the experiment data is most well fitted with $-\alpha$ln($T$), $-\beta$$T^{1/2}$, and $-\gamma$$T^{2}$, the results are plotted in Fig. \ref{figure 2}b, \ref{figure 2}c, and \ref{figure 2}d. $\alpha$, $\beta$ and $\gamma$ are the coefficients of the theoretical fitting line, respectively. It shows that the $R_\mathrm{xx}(T)$ exhibits an apparent crossover from $-\alpha$ln($T$) dependence to $-\beta$$T^{1/2}$ dependence, as shown in the linear region of Fig. \ref{figure 2}b and Fig. \ref{figure 2}c. For temperatures between 30 K and 9 K, $R_\mathrm{xx}(T)$ gradually increases, following a strong $-\alpha$ln($T$) dependence. Obviously, the data show no characteristic transition from a $-\alpha$ln($T$) dependence to a $-\gamma$$T^{2}$ dependence, suggesting that the single channel Kondo effect is not the dominant mechanism in our FGaT system. As the temperature subsequently drops, a strong $-\beta$$T^{1/2}$ dependence emerges and fits well down to 1.8 K, as shown in Fig. \ref{figure 2}c. The $-\beta$$T^{1/2}$ dependence cannot be attributed to electron-electron interaction (EEI), as EEI typically manifests at very low temperatures in the millikelvin range$\cite{yang_unconventional_2019}$. However, the observed $-\beta$$T^{1/2}$ dependence dominates up to 9 K but diminishes below 1 K. This excludes the possibility that the quantum interference effects typically encountered in disorder systems, including EEI and localization, contribute to quantum corrections in the $R_\mathrm{xx}(T)$ behavior$\cite{lee_disordered_1985}$. The observed crossover from a linear $-\alpha$ln($T$) dependence to a $-\beta$$T^{1/2}$ dependence suggests that the $-\beta$$T^{1/2}$ dependence should be attributed to the NFL behavior associated with the orbital 2CK effect$\cite{hewson_kondo_1993}$. To determine the $R_\mathrm{xx}(T)$ at further lower temperatures, we measured $R_{xx}$ from 1.8 K down to 0.36 K for more details. Fig. \ref{figure 2}e illustrates the $-\beta$$T^{1/2}$ trend persisting down to approximately 1 K. Below 1 K, the $R_\mathrm{xx}(T)$ deviates from $-\beta$$T^{1/2}$, and exhibits a clear saturation down to 0.36 K. 

As shown in Fig. \ref{figure 2}, the Kond regime in the FGaT sample starts at $T_\mathrm{0}$ $\sim$ 30 K, $R_\mathrm{xx}$ $\sim$ $-\alpha$ln($T$), then at $T_\mathrm{K}$ around 9.3 K, the $R_\mathrm{xx}(T)$ emerges into the NFL regime, as shown in Fig. \ref{figure 2}c, where $R_\mathrm{xx}(T)$ clearly deviates from the $-\alpha$ln($T$) and can be fitted well with $-\beta$$T^{1/2}$, consistent the orbital 2CK model. At lower temperature below $T_\mathrm{D}$ around 1 K, $R_\mathrm{xx}(T)$ deviates again from the $-\beta$$T^{1/2}$, as shown in Fig. \ref{figure 2}d. One notes that the lowest temperature regime has never been observed simultaneously with the other two regimes previously since the propose of orbital 2CK model, and the $R_\mathrm{xx}(T)$ of the FGaT sample can be well described by the $-\gamma$$T^{2}$ down to 0.36 K as shown in Fig. \ref{figure 2}d, exactly as proposed by Zawadowski$\cite{zawadowski_kondo-like_1980}$. This is the first time that all three regimes of the orbital 2CK model have been observed in a single sample, conclusively demonstrating that the resistance upturn in the FGaT is caused by the orbital 2CK effect. Furthermore, the applied magnetic field only shifts the resistance $R_\mathrm{xx}(T)$, does not affect the critical temperatures of the three-resistive regimes, nor change the coefficients $\alpha$, $\beta$ and $\gamma$ of the three regimes as shown in Fig. \ref{figure 2}b, \ref{figure 2}c and \ref{figure 2}e. This shows that the resistance upturn is nonmagnetic origin, consistent with the orbital 2CK model induced by the TLSs as pseudospin. After the sample was oxidized in atmosphere for 7 days, the longitudinal resistance $R_\mathrm{xx}(T)$ increased from approximately 6 $\Omega$  to 24 $\Omega$ as shown in Fig. \ref{figure 2}d and \ref{figure 2}e, but both the coefficients and $\ce(T_D)$ remain unchanged, this indicates that the origin of the TLSs is not from the surface of the sample, may come from the defects such as grain boundaries, dislocations, twists, and point defects inside the sample as proposed in other systems$\cite{cox_exotic_1998}$.

\begin{figure}[H]
	\centering
	\includegraphics{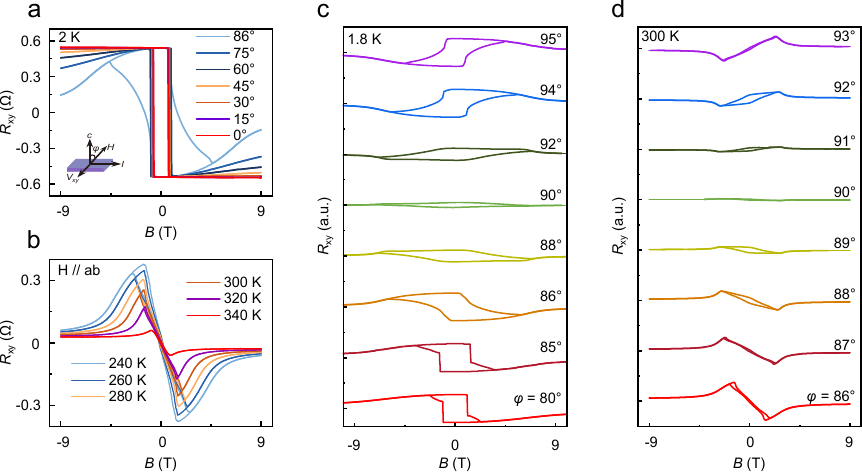}
	\caption{Hall effect in FGaT flakes. (a)$R_{xy}$ vs. $B$, with $H$ rotating from $H$$\parallel$$c$ to $ab$-plane, field up to $\pm$9 T, $T$ = 2 K.(b) $R_{xy}$ vs. $B$ from 240 K to 340 K, with $H$$\parallel$$ab$, field up to $\pm$9 T. (c) $R_{xy}$ vs. $B$ for $\varphi$ from $\varphi = 80^\circ$ to $\varphi = 95^\circ$, field up to $\pm$9 T, $T$ = 1.8 K. (d) $R_{xy}$ vs. $B$ for $\varphi$ from $\varphi = 86^\circ$ to $\varphi = 93^\circ$, field up to $\pm$9 T, $T$ = 300 K.}
	\label{figure 3}
\end{figure}

One potential method to characterize the NFL behavior is the Hall measurement$\cite{ritz_formation_2013}$. We measure Hall effect of the FGaT sample, as shown in Fig. \ref{figure 3}. At 2 K, the abrupt change in Hall signal at the coercive field $H_\mathrm{c}$ suggests rapid domain flipping in FGaT as shown in Fig. \ref{figure 3}a. As $H$ rotates to the $ab$-plane, the $R_{xy}$ decreases and exhibits the expected angular-dependent characteristic. When  $\varphi = 86^\circ$, the square-shaped hysteresis loop disappears, and the magnetic domains undergo a slow flipping process towards the $ab$-plane under the magnetic field. We measure the $R_{xy}$ around the $ab$-plane in detail as shown in Fig. \ref{figure 3}c at 1.8 K. At  $\varphi = 90^\circ$, the hysteresis loop shows a small ’diamond-shaped’ structure with the largest switching field, and the Hall signal switches below and above this angle as expected for a magnetic film with perpendicular anisotropy, indicating $\varphi = 90^\circ$ is near the $ab$-plane. Notably, as temperature increases from 240 K to 320 K, the Hall loop shrinks, two antisymmetric peaks emerge in the Hall signal, as shown in Fig. \ref{figure 3}b. Antisymmetric peaks in Hall signal have been recognized as the topological Hall signal before$\cite{kimbell_challenges_2022}$. We measure the two peaks around the $ab$-plane from $\varphi = 86^\circ$ to $\varphi = 93^\circ$ as shown in Fig. \ref{figure 3}d. The antisymmetric peaks in the Hall signal nearly vanish at approximately  $\varphi = 90^\circ$, very similar to the 1.8 K results. And the two peaks switches below and above this angle, indicating that $H$ is close to alignment with the $ab$-plane near  $\varphi = 90^\circ$. Therefore, the observed Hall effect is simply anomalous Hall of the FGaT sample. We did not find any topological Hall signal in the NFL regime of 1-9 K, and it is not conclusive that the NFL behavior can be detected via the Hall signal$\cite{kimbell_challenges_2022}$.

\begin{figure}[H]
	\centering
	\includegraphics{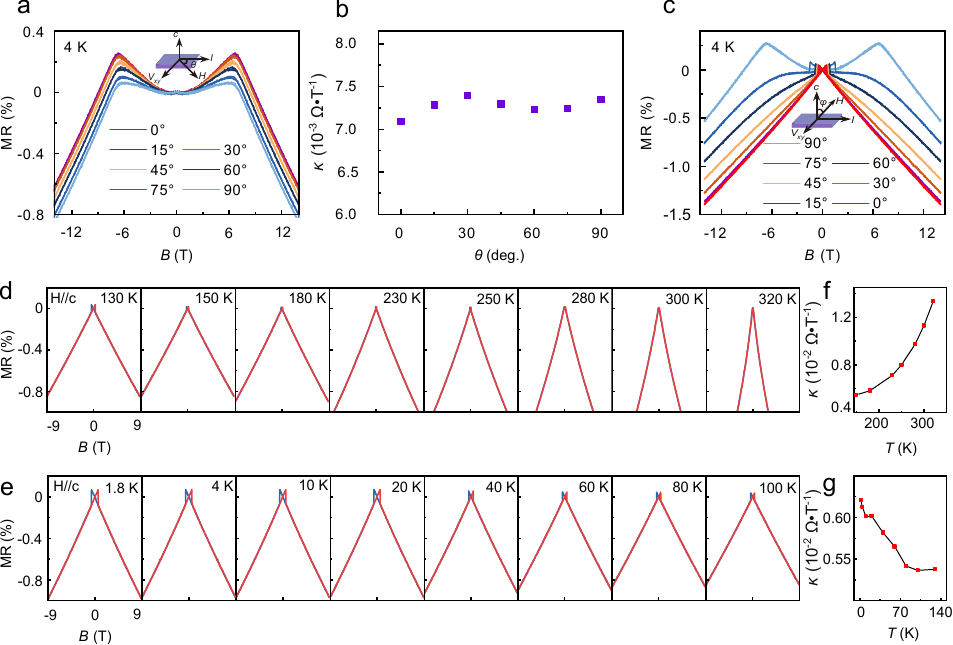}
	\caption{Magnetoresistance measurement of FGaT flakes. (a) MR vs. $B$ in the $ab$-plane up to $\pm$14 T, $T$ = 4 K. (b) Slope magnitude $\kappa$ from LNMR regions plotted against $\theta$ for $B$ > 7 T, slope magnitude $\kappa = |dR_{xx}/dB|$. (c) Angular-dependent MR with $H$ rotating from $ab$-plane to $c$-axis, $T$ = 4 K. (d) MR vs. $B$ along the $c$-axis up to $\pm$9 T, cooling from 320 K to 130 K. (e) MR vs. $B$ along the $c$-axis up to $\pm$9 T, cooling from 100 K to 1.8 K. (f) Slope magnitude $\kappa$ of LNMR plotted against temperature from 320 K to 130 K. (g) Slope magnitude $\kappa$ of LNMR plotted against temperature from 100 K to 1.8 K.}
	\label{figure 4}
\end{figure}

The magnetoresistance of the FeGaT sample is measured in a field up to 14 T in various angle, as shown in Fig. \ref{figure 4}. The magnetic field is initially aligned with the current direction and then rotated within the $ab$-plane. Over the range of $-$14 T to $+$14 T for each angle, MR exhibits two distinct dependencies on the magnetic field. When the magnetic field is less than the in-plane anisotropic field $H_\mathrm{A}$ ($\sim$ $\pm$7 T), a quadratic characteristic with positive MR is observed. This is the expected anisotropic magnetoresistance (AMR) effect$\cite{checkelsky_anomalous_2008}$ since it needs large field to bring magnetization to the field direction, consistent with the Hall measurement in Fig. \ref{figure 3}. Indeed, for field directions from $c$-axis to $ab$-plane, the quadratic feature in small field become smaller and switches at coercive fields, as shown in Fig. \ref{figure 3}c. When the field exceeds about 7 T, the magnetization becomes saturated and aligned with the $ab$-plane, a LNMR appears and does not saturate up to 14 T. Interferingly, the LNMR shows up for field applied in all direction and has almost the same slope, as shown in Fig. \ref{figure 4}a, \ref{figure 4}c. At the same temperature, the slope magnitude $\kappa$ of the LNMR also remain constant as shown in Fig. \ref{figure 4}b for all directions. LNMR has been studied before and has been attributed to spin-magnon scattering in ferromagnets$\cite{raquet_electron-magnon_2002, gil_magnetoresistance_2005}$. The spin-magnon scattering leads to an increase in resistance and the energy of spin waves is linearly suppressed (2$\mu\ce{_B}B$) by an applied field, resulting in the LNMR. Therefore, the slope of the LNMR is a measurement of the spin-magnon scattering and the absolute value of the slope increases for increasing temperature for common ferromagnets$\cite{raquet_electron-magnon_2002}$. 

The LNMR of the FGaT samples are measured from 320 K to 1.8 K, as shown in Fig. \ref{figure 4}d, \ref{figure 4}e. Strangely, the slope of the LNMR does not decreases monotonically, as depicted in Fig. \ref{figure 4}f, \ref{figure 4}g. There is a minimum at about 130 K, which is very different from common ferromagnets such as Fe, Co, Ni$\cite{raquet_electron-magnon_2002}$. One notes that an increasement of the slope indicates a stronger spin-magnon scattering at temperature below 130 K. This counterintuitive phenomenon suggests the presence of misalignment in moment in this temperature range. Recently, it has been observed that there exist topological spin textures such as Skyrmion bubbles in the FGaT material between 100 K and 200 K$\cite{saha_high-temperature_2024}$, which may be the origin of the observed increases of LNMR slope here. Notably, the Hall signal in Fig. \ref{figure 3} of the same sample does not show any anomaly. The spin misalignment induced by the spin texture effectively increases spin-magnon scattering. 

Our work provides the first demonstration for the complete orbital 2CK model with three-resistive regimes. The Kondo regime starts at 30 K and the NFL temperature range is from 9 K to 1 K. In 1CK effect, electrons are scattered by impurities causing the first regime with resistance following $-$ln($T$) relevance while it becomes $-T^{2}$ relevance for decreasing temperature when the impurity is fully compensated by the electron spin. The NFL behavior only appears when two electrons compete to compensate the impurity where many exotic phenomena occur which are important to explore high temperature superconductor, Majorana, and other effects that related to quantum criticality. The vdWs ferromagnet FGaT provides a material platform to study NFL behavior in a temperature range of 8 K with two well-defined FL regimes.

In summary, we have observed the full three-resistive regimes in a single FGaT sample, providing unambiguous evidence for the orbital 2CK effect. The transport result indicates that the orbital 2CK effect originates from nonmagnetic TLSs. No topological hall effect is observed in Hall measurement excludes the possibility of magnetism-related NFL behavior below 30 K. The non-saturated LNMR up to 14 T reveals the spin-magnon scattering in FGaT sample. However, the slope of LNMR decreases non-monotonically around 130 K due to the emergence of spin texture in FGaT indicating the slope of LNMR may be a good tool to study topological spin textures in ferromagnets. Our research advances the comprehension of the orbital two-channel Kondo effect and establishes an experimental framework to study non-Fermi liquid behavior with well-defined Fermi liquid boundaries.

\section{Experimental}
\subsection{Crystal growth}

The $\ce{Fe_3GaTe_2}$ single crystals were grown using the self-flux method. Fe powder (99.99\%), Ga ingots (99.98\%), and Te powder (99.99\%) were homogeneously mixed in a molar ratio of 1:1:2 within a glove box and subsequently sealed in a vacuum-sealed quartz tube. The mixture was rapidly heated to 1000 °C within  1 hour and maintained at 1000 °C for 24 hours. Subsequently, the temperature was rapidly decreased to 780 °C within 1 hour and held at 780 °C for 100 hours. Finally, the sample was subjected to a temperature-controlled centrifugation to separate the $\ce{Fe_3GaTe_2}$ crystals from the flux. 

\subsection{Structure characterizations}
X-ray diffraction measurements were conducted on the \ce{Fe_3GaTe_2} sample with a Rigaku Smartlab 3K, yielding the (00l) diffraction peaks. Subsequently, the elemental composition of the \ce{Fe_3GaTe_2} samples was qualitatively measured and elemental mapping images were acquired using an energy dispersive spectrometer (EDS). A single-crystal X-ray diffractometer (Bruker D8 VENTURE) was utilized to determine the precise elemental composition and ratios, and diffraction images of the \ce{Fe_3GaTe_2} single-crystal samples were obtained.

\subsection{Device fabrication and transport measurement}

The high quality single crystal \ce{Fe_3GaTe_2} was first mechanically exfoliated with silicon-free blue tape and transfered onto \ce{SiO_2}/Si substrate. The standard hall bar electrode was patterned by laser direct writing machine (DWL 66+) and then coating with Ti/Au (5 nm/50 nm) using electron beam evaporation coating system (JEB-2) . All transport measurements were carried out in a physical property measurement system (PPMS DynaCool)  with a base temperature of 1.8 K and a magnetic field of up to 14 T. Additionally, the Helium-3 cryostat accessory for PPMS enables measurement temperatures down to 500 mK.

\begin{acknowledgement}

The authors acknowledge support from the Key-Area Research and Development Program of Guangdong Province (Grants Nos. 2020B0303050001, 2021B0101300001), the National Natural Science Foundation of China (Grants Nos. 11974158, 11804402, 22205091, 12204221), the Guangdong Basic and Applied Basic Research Foundation (Grant No. 2022A1515012283, 2024A1515030118)

\end{acknowledgement}

\section{Author Contributions}
T.Y.C. supervised the project. C.H.B. fabricated and characterized the thin-layer devices with support from J.F.S., X.L.Y., L.X.L., and Z.Y.L. X.L.Y. and X.M.M. provided the $\ce{Fe_3GaTe_2}$ single crystals. S.G. conducted the SCXRD measurements. C.H.B., and T.Y.C. analyzed the results and prepared the manuscript with input from others.

\bibliography{achemso-demo}

\end{document}